\documentclass[conference]{IEEEtran}
\usepackage{cite}
\usepackage{amsmath,amssymb,amsfonts}
\usepackage{algorithmic}
\usepackage{graphicx}
\usepackage{textcomp}
\usepackage{xcolor}
\usepackage{tikz}
\usepackage{pgfplots}
\usepackage[mode=buildnew]{standalone}
\pgfplotsset{compat=1.18}
\usepackage{gensymb}
\usepackage{booktabs}
\usetikzlibrary{fit,arrows,positioning,arrows.meta}
\usetikzlibrary{patterns}
\usetikzlibrary{shapes.arrows}
\usepackage{algorithm}
\usepackage{algorithmic}
\usepackage{hyperref}
\usepackage{fancyhdr}
\hypersetup{          
    colorlinks=true,
    linkcolor=black,
    filecolor=magenta,      
    urlcolor=cyan,
    citecolor=teal!80,
    pdftitle={},
    }

\fancypagestyle{firstpage}{
    \fancyhf{} 
    \fancyfoot[L]{979-8-3503-5685-4/24/\$31.00 \copyright2024 IEEE}

}

\newcommand{\subparagraph}{}
\usepackage{titlesec}
 \titlespacing*{\section}{0pt}{3pt plus 1pt minus 1pt}{1pt plus 1pt minus 2pt}
 \titlespacing{\subsection}{0pt}{0pt plus 0pt minus 1pt}{0pt plus 0pt minus 1pt}
 \titlespacing{\subsubsection}{0pt}{1pt plus 0pt minus 0pt}{1pt}

\def\BibTeX{{\rm B\kern-.05em{\sc i\kern-.025em b}\kern-.08em
    T\kern-.1667em\lower.7ex\hbox{E}\kern-.125emX}}
\begin{document}
\title{USRP-Based Single Anchor Positioning: AoA with 5G Uplink Signals, and UWB Ranging}

\author{\IEEEauthorblockN{Thodoris Spanos$^{1,2}$, 
Fran Fabra$^3$, Jos\'e A. L\'opez-Salcedo$^3$, Gonzalo Seco-Granados$^3$, \\Nikos Kanistras$^2$, {Ivan Lapin$^4$}, Vassilis Paliouras$^1$
}\vspace{0.1cm}
\IEEEauthorblockA{$^1$Dept. of Electrical and Computer Engineering, University of Patras, Greece\\ $^2$Loctio P.C., Greece\\
$^3$Dept. of Telecommunication and Systems Eng., Universitat Aut\`onoma de Barcelona (UAB), Spain\\
 $^4$Radio Navigation Systems and Techniques Section, European Space Agency, The Netherlands}
}

\maketitle
\thispagestyle{firstpage}
\begin{abstract}
This paper presents a novel testbed designed for 5th-Generation (5G) positioning using Universal Software Radio Peripherals (USRPs). The testbed integrates multiple units: an Operation Unit for test management, a User Unit equipped with an Ettus E312 USRP, and a Station Unit featuring an Ettus N310 USRP equipped with a three-element Uniform Linear Array for Angle of Arrival estimation. Alongside ultra wide-band ranging, the testbed estimates the user's position relative to the base station. Signal processing algorithms are executed in a dedicated processing unit. Key challenges addressed include phase misalignment between RX channel pairs due to different Local Oscillators in the Ettus N310, necessitating real-time calibration for precise signal alignment. High sampling rates (up to 61.44 MSps) result in large IQ sample files, managed efficiently using a snapshot technique to optimize storage without compromising testbed positioning capabilities. The testbed synchronizes angular measurements with ranging estimates allowing consistent performance evaluation for real-life cases of dynamic users (e.g. pedestrian). Experimental results demonstrate the testbed's effectiveness in achieving accurate pedestrian user localization.
\end{abstract}

\begin{IEEEkeywords}
testbed, AoA, 5G, 6G, Positioning, UWB, Single, Anchor, USRP
\end{IEEEkeywords}
\vspace*{-10pt}
\section{Introduction}
Although the 5th Generation (5G) technology is commercially considered still in its early stages, it has matured significantly in recent years. Specifically in localization, the ability to leverage large bandwidths, high frequencies, and Ultra-Dense Networks has enabled communication systems to autonomously estimate user positions and map propagation environments \cite{9693225}. 
Continued advancements in positioning capabilities are anticipated with the rise of 5G technologies, which are essential for applications such as indoor navigation and real-time drone tracking \cite{10041419}. Current 5G networks rely on proximity information, fingerprints, or single-value estimates like time and power for localization. Techniques such as angle of arrival (AoA), downlink time-difference-of-arrival and uplink time-difference-of-arrival, are instrumental in achieving higher localization accuracy \cite{10127621}. 

Positioning using a single node or station is an alternative method to determine the user position by relying on angular and distance measurements, making it particularly suitable for environments with limited availability of the Global Navigation Satellite System (GNSS), such as indoors or deep urban canyons. As we move towards 6G and FR3, positioning using a single node with large antenna arrays in existing 5G systems has gained renewed research interest, laying the groundwork for future advancements in localization technologies. Simulation results utilizing various techniques show high accuracy even in challenging conditions. Li \emph{et al.} use the Multiple Signal Classification (MUSIC) algorithm to estimate time-of-flight and AoA, providing centimeter-level accuracy \cite{9348766}. Sun \emph{et al.} explored 3D positioning by jointly estimating Time of Arrival (ToA)-AoA with a Uniform Rectangular Array \cite{3D}. A high-precision indoor positioning method using a single Base-Station (BS) and 5G signals is presented by Liu \emph{et al.} \cite{9441655}. 
Additionally Xie \emph{et al.} propose a scattering area model with specific spatial layout information for outdoor single-station positioning in an NLOS environment \cite{9886342}. 

However, real-world experimentation---especially with 5G signals---remains limited. Given the diverse range of use cases, making direct comparisons between theoretical and experimental results, or even among different experimental outcomes, is nearly impossible. A testbed is deployed using NI Software-Defined Radios (SDRs), measuring the AoA for different distances using MUSIC and Estimation of Signal Parameters via Rotational Invariance Techniques (ESPRIT) by Rares \emph{et al.} \cite{8514133}. Blanco \emph{et al.} evaluate a Long Term Evolution (LTE)-based positioning testbed in 2D space using ToA estimation \cite{LTEtb}. The authors use LTE Sounding Reference Signals (SRS) and AoA measurements utilizing a bladeRF SDR as their user and an Ettus X310 as their BS. 
A 5G system using OpenAirInterface was built by Li \emph{et al.} implementing the Enhanced Cell-ID positioning method with ToA and AoA estimation, along with an error reduction method, meeting indoor commercial requirements \cite{9880817}. Ge \emph{et al.} report initial results on single-BS vehicular positioning using downlink 5G mmWave signals, highlighting gaps between theoretical assumptions and real-world behavior \cite{9841230}. 

In this paper, we explore advancements in single-node positioning using a novel testbed setup based on Universal Software Radio Peripherals (USRPs) utilizing the SRS in the 5G New Radio Uplink. 
Using Commercial-Off-The-Shelf equipment and advanced signal processing algorithms, our research aims to showcase the feasibility and performance of single-node positioning systems in practical applications, paving the way for enhanced localization capabilities in diverse operational environments. The remainder of the paper is organized as follows: Section~\ref{testbed_setup} presents the testbed setup. Section~\ref{testbed_operation} describes the key testbed functionalities. Section~\ref{field_trials} discusses the field trial results. Section~\ref{conclusions} concludes the paper.
\section{Testbed Setup}\label{testbed_setup}
\subsection{Overview}
The primary objective of the testbed is to utilize two Ettus USRPs---an E312 as the transmitter and an N310 as the receiver---for AoA estimation using SRS in the uplink. By integrating angular estimation with ultra-wideband (UWB) measurements for range and barometer readings for altitude, the testbed achieves precise position estimation relative to the known position of the base station. The testbed operates in the 2.4 GHz, 3.5 GHz, and 5.8 GHz frequency bands. Additionally, it supports eight 5G waveforms for two different bandwidths (20 MHz and 50 MHz) of numerologies 1 and 2. The supported configurations are shown in Table~\ref{Waveform Configurations}.
\begin{table}[tb]
\centering
\caption{5G Waveform Configurations\vskip9pt}
\label{Waveform Configurations}
\resizebox{0.48\textwidth}{!}{%
\footnotesize
\begin{tabular}{lcccccl}\toprule
Conf. & Use Case & Frequency & Subcarrier & Bandwidth \\
& & Band (GHz) & Spacing (kHz) & (MHz)\\
\midrule
I & Static, Pedestrian & 2.4 & 30 & 20 \\ 
II & Static, Pedestrian & 2.4 & 30 & 50 \\
III & Static, Pedestrian & 3.5 & 30 & 20 \\ 
IV & Static, Pedestrian & 3.5 & 30 & 50 \\
V & Vehicular & 3.5 & 60 & 20 \\ 
VI & Vehicular & 3.5 & 60 & 50 \\
VII & Pedestrian, Vehicular & 5.8 & 60 & 20 \\ 
VIII & Pedestrian, Vehicular & 5.8 & 60 & 50 \\
\bottomrule
\end{tabular}
}
\label{Configuration}
\end{table}

The testbed consists of four distinct units, each with a pivotal role in its operation: the Operation Unit (OU), the User Unit (UU), the Station Unit (SU), and the Processing Unit (PU). The OU serves as the interface for the operator, enabling test initiation and report viewing. The UU includes the Ettus E312 connected to a laptop, the UWB tag and a barometer, facilitating user-side operations. Meanwhile, the SU comprises the Ettus N310 connected to a host PC, alongside the UWB anchor module and a barometer, serving as the base station. Lastly, the PU is dedicated to executing the signal processing algorithms supported by the testbed. While each unit can be hosted on a different server to support remote operation or faster processing, the OU, SU and PU have been deployed on a single host PC for efficiency.
\subsection{SRS Configuration}
The SRS is specifically designated for positioning in the uplink by the 5G standard due to its high autocorrelation properties and flexible configuration \cite{3gppsrs}. In the proposed testbed, it is transmitted over the entire signal bandwidth, periodically in every slot, and is contained in the first four OFDM symbols of the slot. The SRS pilots are mapped to the physical resources according to a comb-like pattern every $K_{TC} {=} 2$ subcarriers, which provides the highest density of SRS pilots in the frequency domain.
\subsection{Equipment}
\begin{figure*}
    \centering
    \includegraphics[width=1\textwidth,height=0.21\textheight]{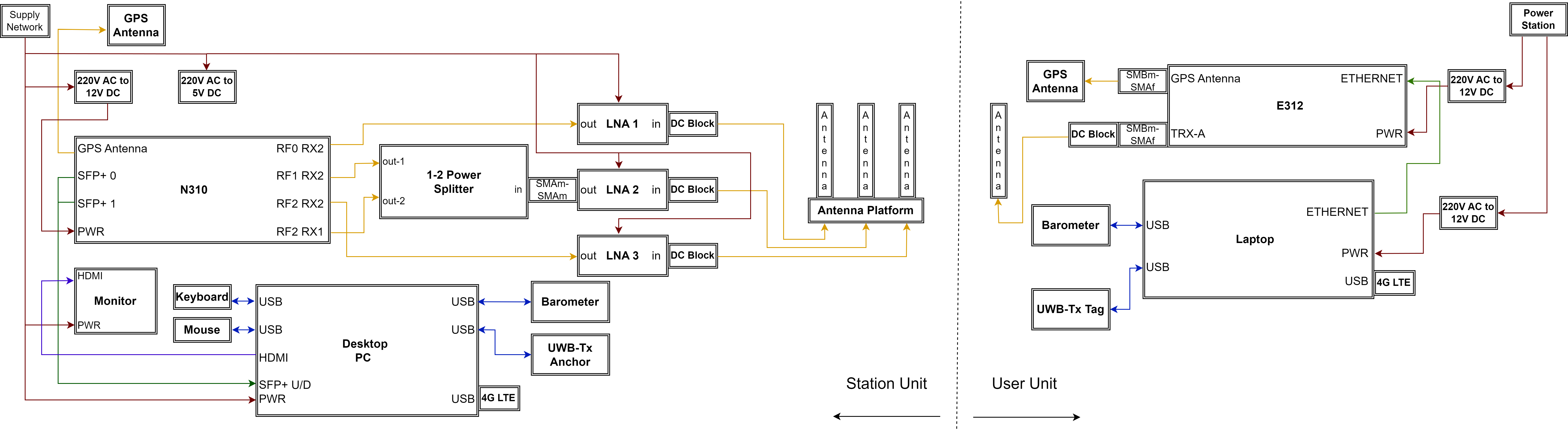}
    \caption{Block diagram presenting the testbed architecture. Operation and Processing Units are hosted in the Station Unit PC.}
    \label{block_setup_real}
\end{figure*}

In this section, the core testbed components are analysed. In addition, a block diagram of the testbed setup containing all the installed components is presented in Fig.~\ref{block_setup_real}.
\subsubsection{Ettus E312}
The Ettus USRP E312 is a portable SDR platform designed for field deployment. It features a flexible 2$\times$2 transceiver, providing up to 56 MHz of instantaneous bandwidth and covering frequencies from 70 MHz to 6 GHz. Additionally, the Ettus E312 is equipped with a GPS receiver.
\subsubsection{Ettus N310}
The Ettus USRP N310 is a high-performance SDR equipped with a direct conversion 4$\times$4 transceiver architecture. It supports up to 100 MHz of instantaneous bandwidth per channel and covers frequencies from 10 MHz to 6 GHz. Furthermore an integrated GPS receiver is also included.
\subsubsection{Decawave EVK1000}
The UWB Decawave EVK1000 evaluation kit is designed for ranging and localization applications. It performs two-way ranging between two DW1000 transceivers and estimates the distance between them based on the time-of-flight calculations. 
\subsubsection{Barometer}
A pair of barometers (Grove - Barometer operating through Arduino \cite{grove_barometer}) is installed in both the UU and the SU to measure the altitude difference. However, we were unable to extract accurate measurements; even at similar altitudes, the height difference readings are unstable. Consequently, the barometer results are omitted from the final analysis.
\subsubsection{Uniform Linear Array}
Ettus N310 consists of four Rx channels. Three of them are used for AoA estimation, fed by a three-element ULA to capture the transmitted signal, as detailed in Section~\ref{calibration_section}. The distance between consecutive ULA elements is $\lambda/2$.
\section{Testbed Operation}\label{testbed_operation}
\subsection{Ettus N310 Calibration Procedure}\label{calibration_section}
Although the Ettus N310 consists of four Rx channels, which would theoretically allow for a four-element ULA, only three of them can be utilized effectively without the use of an external clock or a phase offset compensation procedure after each initialization \cite{Xhafa2023}, \cite{me}. This is due to the architecture of the Ettus N310, which includes two daughterboards, each associated with a pair of Rx channels. Initially, both channels within each pair exhibit phase misalignment, necessitating a calibration procedure. This calibration involves using a 1\mbox{-}4 splitter to feed all four channels the same tone signal, ensuring both channels of each pair receive identical inputs. The phase difference between the two channels of each pair is then computed by cross-correlating the received signals, as the only variation between them is the phase offset. The calculated phase differences are stored and applied for phase offset correction to one of the two Rx channels in each pair during signal processing, to align them in phase.

Moreover, the Ettus N310's daughterboards use different Local Oscillators (LOs) for their respective Rx channel pairs, leading to random phase variations between runs. To address these phase offsets, a real-time calibration process is introduced. During normal testbed operations, a common signal is injected into one channel of each pair via a 1-2 splitter. This setup allows the differential phase caused by the different LOs to be measured, albeit limiting the ULA utilization to three channels. The inherent phase difference is then compensated in real-time, ensuring phase alignment across the channels. 
\subsection{Snapshot Capture}
Given the high sampling rates supported by the testbed, i.e., 30.72 MSps and 61.44 MSps, a 10 Gigabit Ethernet card is used to transfer data to the host PC at a high data rate, utilizing two SFP+ cables. As a result of the used sampling rate, the generated IQ sample files are very large, reaching hundreds of GB for just a few minutes of testing. However, since processing just a few milliseconds of the received signal can yield consistent AoA estimation, a snapshot-capturing approach has been preferred, significantly reducing the storage and processing requirements without compromising the positioning accuracy of the testbed.
\subsection{Software}
To facilitate real-life field tests, automated control of the testbed units is necessary. Dedicated Python software has been developed to manage the testbed control, handling various operations such as configuring the USRPs, managing transmission start/finish cycles, overseeing data capture processes, and logging measurements from the peripherals (e.g., UWB, barometers, integrated GPS receivers). 

\begin{figure}
    \centering    \includegraphics[width=0.48\textwidth]{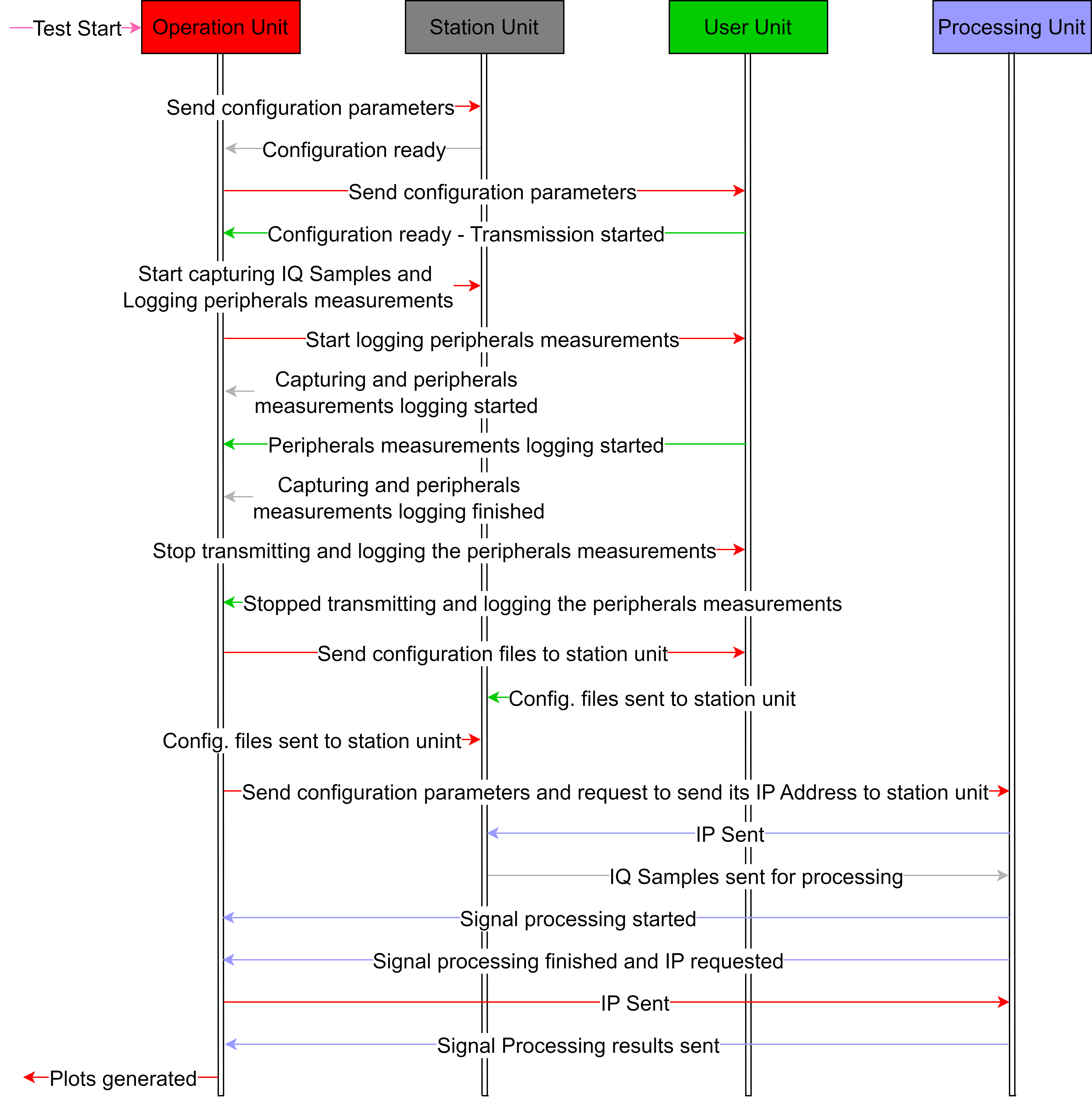}
    \caption{Communication between the different units of the testbed.}
    \label{communication}
\end{figure}

For inter-unit communication within the testbed, the Message Queuing Telemetry Transport (MQTT) protocol has been implemented. This protocol efficiently manages communication between the OU, UU, SU, and PU. A remote MQTT broker is deployed on a separate server, independent of the testbed units, facilitating message exchange crucial for testbed control. Each unit within the testbed operates an MQTT client, enabling it to function as publisher and/or subscriber. This setup ensures seamless transmission of commands and data between units, supporting coordinated operations such as test initiation, status updates, and data synchronization. Moreover, the testbed supports scalability, allowing the integration of additional user units. The testbed communication between different units is depicted in Fig.~\ref{communication}.
\subsection{Signal Processing}\label{sp}
After capturing the IQ samples, the data undergoes processing, including signal processing to compute an AoA estimation per second and synchronizing angular measurements with ranging measurements and ground truth. Signal processing performed by the testbed, includes eight key actions for each captured snapshot: 
\begin{enumerate}
    \item Loading a number of IQ samples corresponding to the snapshot length.
    \item Compensating the phase offset of Ettus N310 channels.
    \item Performing timing synchronization by cross-correlation with a stored replica of the transmitted signal \cite{ElHajjar2007}.
    \item Performing frequency offset correction by correlating the captured signal with the cyclic prefix \cite{cfo}.
    \item Estimating channel order using the Akaike criterion \cite{Akaike}.
    \item Estimating the AoA with one of: MUSIC \cite{MUSIC}, ESPRIT \cite{ESPRIT}, 2D ESPRIT \cite{2DESPRIT}, ROOT-MUSIC \cite{ROOT-MUSIC}, or Minimum Variance Distortionless Response \cite{MVDR} algorithms.
    \item Using the Linear Constraint Minimum Variance beamformer \cite{lcmv} to select an angular estimation when the estimated channel order is greater than one.
    \item Computing the Signal to Interference plus Noise Ratio by measuring the power in the empty subcarriers (noise) and subtracting the mean from the power measured in the utilized subcarriers (signal plus noise).
\end{enumerate}
\section{Field Trials}\label{field_trials}
\subsection{Overview}
\begin{figure}[b]
    \centering
    \includegraphics[width=0.48\textwidth]{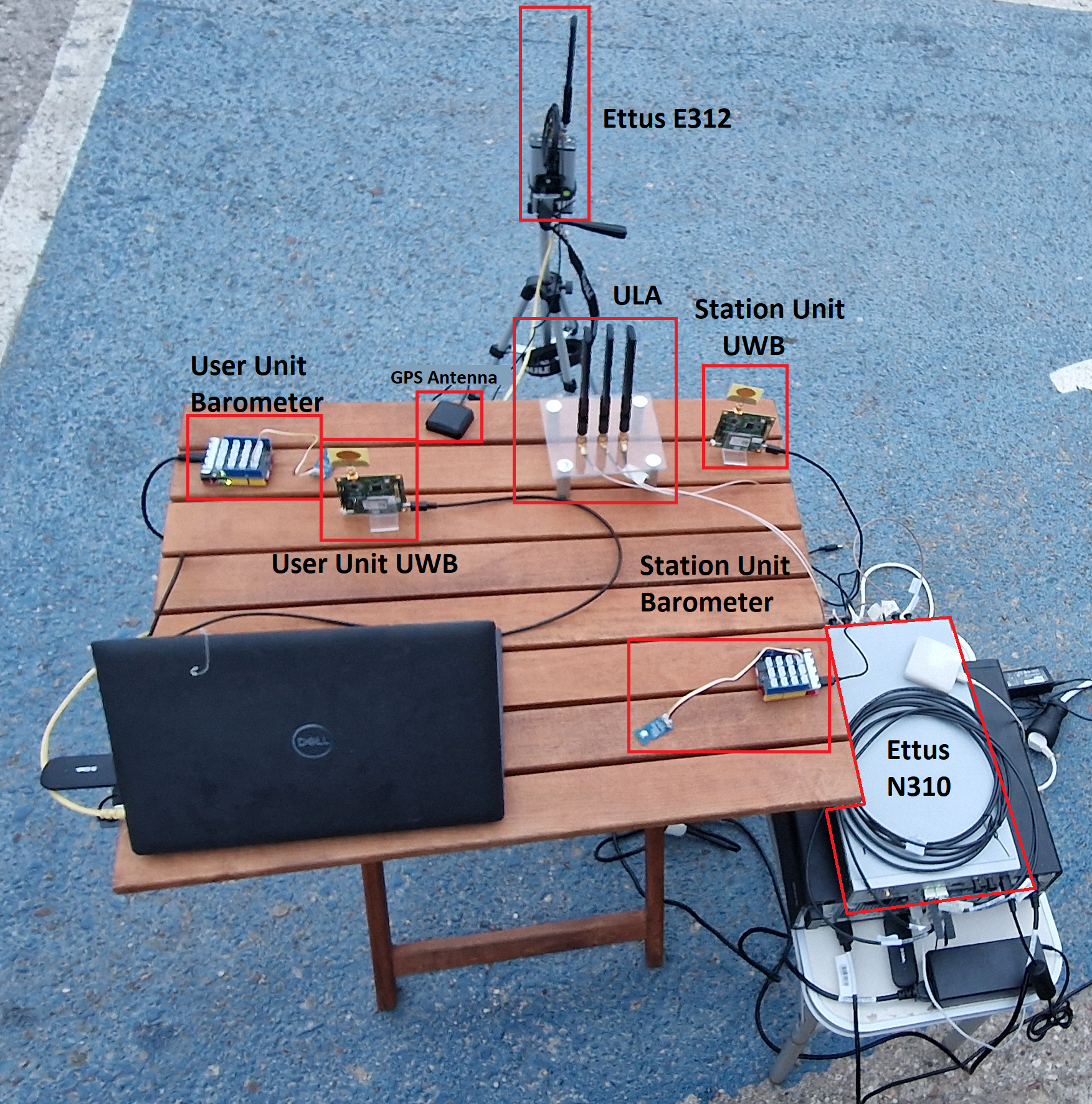}
    \caption{Equipment for Field Trials.}
    \label{singpos_equipment_field_trials}
\end{figure}
The equipment setup on the field is shown in Fig.~\ref{singpos_equipment_field_trials}. A series of validation tests ensured effective testbed operation across three scenarios, static \cite{me}, pedestrian, vehicular. This paper focuses on the pedestrian case, using the Industrial, Scientific and Medical (ISM) 2.4~GHz and 5.8~GHz bands using configurations II and VIII from Table~\ref{Waveform Configurations}. The algorithms in Section~\ref{sp}, previously evaluated in simulations for execution time and performance, identified ESPRIT as the preferred choice for the field trials. The test trajectory, illustrated in Fig.~\ref{trajectory}, starts at point A1 (15 m from the station unit antenna), proceeds through points A2, T3, T2, T1, and returns to A1, with a maximum distance of 90 m at T2. The pedestrian user moves at 4 km/h, remaining stationary for 30 s at each point.
\subsection{Ground Truth}
The landmarks along the trajectory were precisely measured relative to the base station using a laser measure, providing known reference points. The high accuracy of the UWB device enabled reliable extraction of timestamps at the stationary landmarks, where the user remained for 30 s, ensuring precise temporal synchronization. By interpolating between these stationary points, an accurate ground truth trajectory was established to reliably validate the testbed's performance.
\subsection{Results}
\begin{figure}
\begin{tabular}{cc}
\centering
\begin{tikzpicture}
\begin{axis}[legend style={at={(1,0.0)},anchor=south east},legend style={nodes={scale=1, transform shape}},xlabel=Position Error (m), ylabel={\shortstack{Cumulative Density Function}}, xmin=0, xmax=30, ymin=0, ymax=1,ytick={0,0.1,0.2,0.3,0.4,0.5,0.6,0.7,0.8,0.9,1},
    yticklabels={0,0.1,0.2,0.3,0.4,0.5,0.6,0.7,0.8,0.9,1},    grid=both,             clip=false,
    unbounded coords=jump]
    \addplot[red, mark=None] table[x index = 1, y index = 0]{05XX/0510/Data/cdf_vld_test_0510_SID2_F2p451GHz_TxG89_RxG75_TD300s_SL9ms.tex};
    \addlegendentry{Position Error CDF}
\end{axis}
\end{tikzpicture} &

\centering
\begin{tikzpicture}
\begin{axis}[legend style={at={(1,0.0)},anchor=south east},legend style={nodes={scale=0.7, transform shape}},xlabel=Position Error (m), ylabel={\shortstack{Cumulative Density Function}}, xmin=0, xmax=30, ymin=0, ymax=1,ytick={0,0.1,0.2,0.3,0.4,0.5,0.6,0.7,0.8,0.9,1},
    yticklabels={0,0.1,0.2,0.3,0.4,0.5,0.6,0.7,0.8,0.9,1},    grid=both,             clip=false,
    unbounded coords=jump]
    \addplot[red, mark=None] table[x index = 1, y index = 0]{06XX/0610/Data/cdf_vld_test_0610_SID8_F5p8GHz_TxG89_RxG75_TD300s_SL9ms.tex};
    \addlegendentry{Position Error CDF}
\end{axis}
\end{tikzpicture}
\end{tabular}
    \caption{Position Error CDF for 2.4 GHz (left) and 5.8 GHz (right).}
    \label{CDF}
\end{figure}
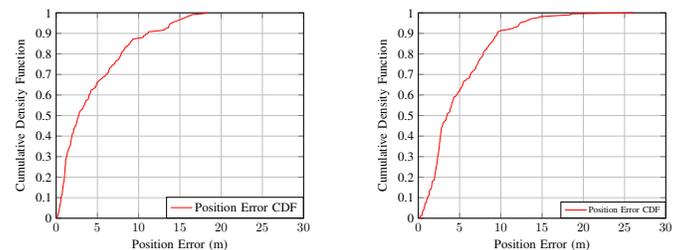

The estimated trajectory compared to the ground truth is depicted in Fig.~\ref{position}, and the position error Cumulative Distribution Function (CDF) is shown in Fig.~\ref{CDF}, with position measurements taken at a frequency of 1 per second. The testbed performs better at shorter distances (up to 40 m) than at longer distances (greater than 40 m) and achieves higher accuracy when the angle is closer to 0\degree{} relative to the ULA center compared to its edges. Additionally, beyond 40 m, the UWB device becomes less reliable, occasionally failing to provide a position estimate every second. This reliability further decreases with distance, resulting in fewer position marks at longer ranges. Nevertheless, the trajectory is accurately followed throughout the entire test duration. Furthermore, the testbed exhibits slightly better performance at 5.8~GHz compared to 2.4~GHz. At 5.8~GHz, the CDF indicates a position error of less than 10 m in 90\% of cases, compared to 88\% at 2.4 GHz. For distances below 40 m, the CDF shows that 90\% of position errors are under 4.6 m at 2.4~GHz and under 3.4 m at 5.8~GHz, demonstrating improved accuracy.
\begin{figure}
    \centering
    \includegraphics[width=0.48\textwidth]{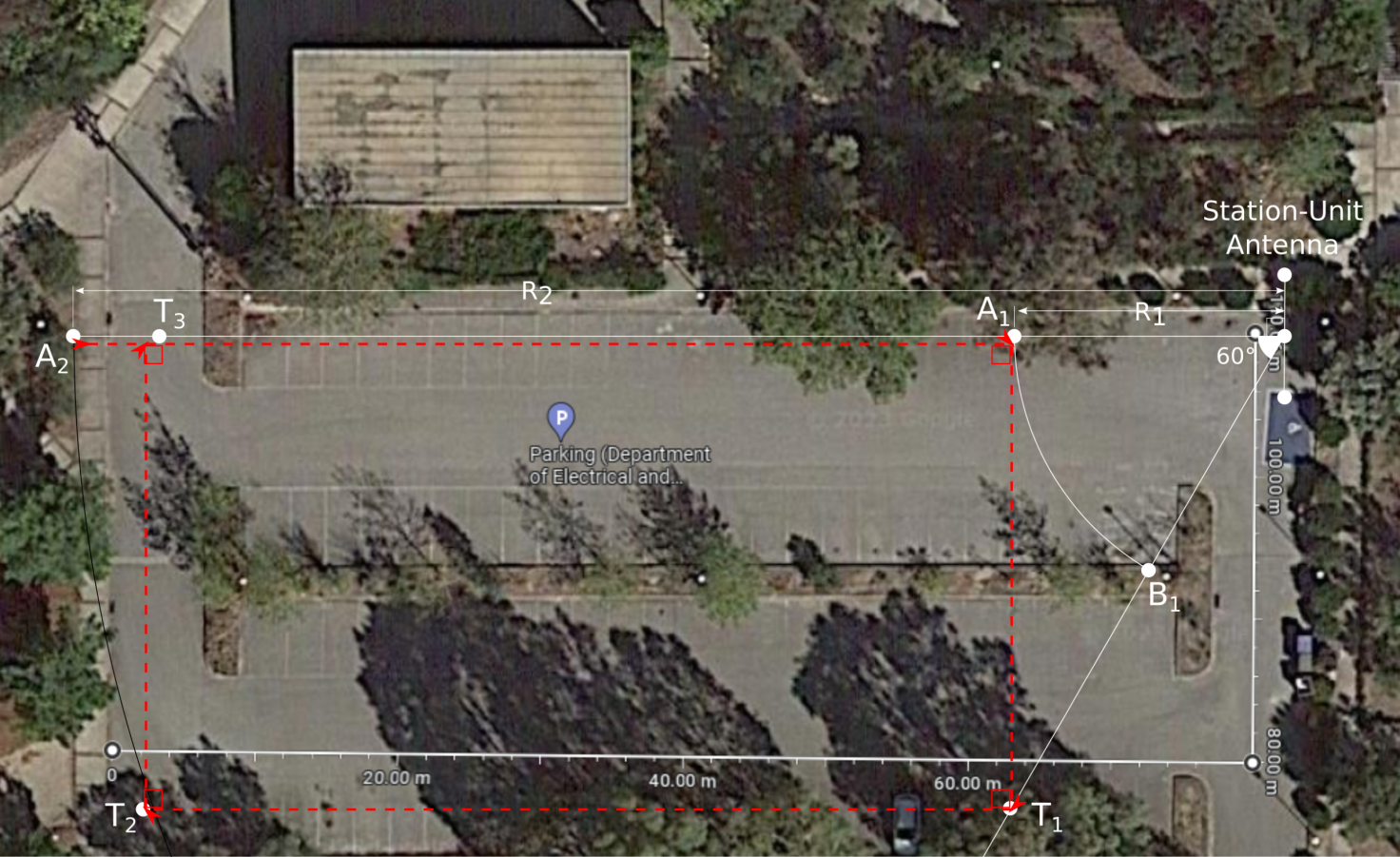}
    \caption{Pedestrian tests trajectory.}
    \label{trajectory}
\end{figure}
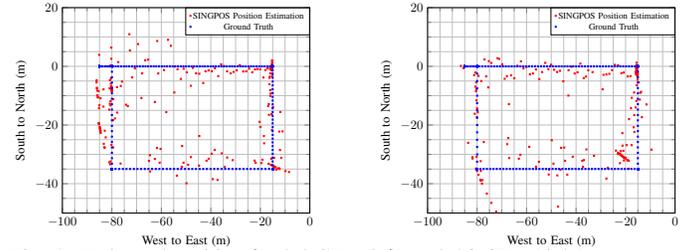
\begin{figure}
\begin{tabular}{cc}
\centering
\begin{tikzpicture}
\begin{axis}[legend style={at={(1,1)},anchor=north east},legend style={nodes={scale=0.7, transform shape}},xlabel=West to East (m), ylabel={\shortstack{South to North (m)}}, xmin=-100, xmax=0, ymin=-50, ymax=20,
        grid=both,   minor x tick num=3,   minor y tick num=3,       clip=false,
    unbounded coords=jump]
    \addplot[red,mark=x, only marks, mark size=1pt] table[x index = 11, y index = 12]{05XX/0510/Data/vld_test_0510_SID2_F2p451GHz_TxG89_RxG75_TD300s_SL9ms.tex};
    \addplot[blue,mark=x, only marks, mark size=1pt] table[x index = 13, y index = 14]{05XX/0510/Data/vld_test_0510_SID2_F2p451GHz_TxG89_RxG75_TD300s_SL9ms.tex};
    \addlegendentry{SINGPOS Position Estimation}
    \addlegendentry{Ground Truth}
\end{axis}
\end{tikzpicture} &

\centering
\begin{tikzpicture}
\begin{axis}[legend style={at={(1,1)},anchor=north east},legend style={nodes={scale=0.7, transform shape}},xlabel=West to East (m), ylabel={\shortstack{South to North (m)}}, xmin=-100, xmax=0, ymin=-50, ymax=20,
        grid=both,   minor x tick num=3,   minor y tick num=3,       clip=false,
    unbounded coords=jump]
    \addplot[red,mark=x, only marks, mark size=1pt] table[x index = 11, y index = 12]{06XX/0610/Data/vld_test_0610_SID8_F5p8GHz_TxG89_RxG75_TD300s_SL9ms.tex};
    \addplot[blue,mark=x, only marks, mark size=1pt] table[x index = 13, y index = 14]{06XX/0610/Data/vld_test_0610_SID8_F5p8GHz_TxG89_RxG75_TD300s_SL9ms.tex};
    \addlegendentry{SINGPOS Position Estimation}
    \addlegendentry{Ground Truth}
\end{axis}
\end{tikzpicture}
\end{tabular}
    \caption{Estimated position for 2.4 GHz (left) and 5.8 GHz (right).}
    \label{position}
\end{figure}

\vspace*{-5pt}
\section{Conclusions}\label{conclusions}
\vspace*{-5pt}
In this paper, a single-anchor positioning testbed using USRPs for a pedestrian user use case has been presented. The testbed integrates multiple components: an Ettus E312 as the transmitter and an Ettus N310, equipped with a three-element ULA for AoA estimation, as the receiver. Range is determined using an UWB pair, enabling accurate user position estimation. Experimental results indicate sub-10 meter accuracy in 88\% and 90\% of the cases for the ISM 2.4 GHz and 5.8 GHz bands respectively, highlighting the testbed's quality performance.
\vspace*{-5pt}
\section{Acknowledgements}
\vspace*{-4pt}
The undertaken efforts were conducted within the framework of the Single Node Positioning Testbed (SINGPOS) project funded by the European Space Agency (ESA).
\IEEEtriggeratref{1}
\IEEEtriggercmd{\enlargethispage{-9.58cm}}
\bibliographystyle{IEEEtran}
\bibliography{IEEEabrv,navitec}

\begin{thebibliography}{10}
\providecommand{\url}[1]{#1}
\csname url@samestyle\endcsname
\providecommand{\newblock}{\relax}
\providecommand{\bibinfo}[2]{#2}
\providecommand{\BIBentrySTDinterwordspacing}{\spaceskip=0pt\relax}
\providecommand{\BIBentryALTinterwordstretchfactor}{4}
\providecommand{\BIBentryALTinterwordspacing}{\spaceskip=\fontdimen2\font plus
\BIBentryALTinterwordstretchfactor\fontdimen3\font minus \fontdimen4\font\relax}
\providecommand{\BIBforeignlanguage}[2]{{%
\expandafter\ifx\csname l@#1\endcsname\relax
\typeout{** WARNING: IEEEtran.bst: No hyphenation pattern has been}%
\typeout{** loaded for the language `#1'. Using the pattern for}%
\typeout{** the default language instead.}%
\else
\language=\csname l@#1\endcsname
\fi
#2}}
\providecommand{\BIBdecl}{\relax}
\BIBdecl

\bibitem{9693225}
C.~Baquero~Barneto, E.~Rastorgueva-Foi, M.~F. Keskin, T.~Riihonen, M.~Turunen, J.~Talvitie, H.~Wymeersch, and M.~Valkama, ``{{Millimeter-Wave Mobile Sensing and Environment Mapping: Models, Algorithms and Validation}},'' \emph{IEEE Transactions on Vehicular Technology}, vol.~71, no.~4, pp. 3900--3916, 2022.

\bibitem{10041419}
A.~K. Dutta and M.~Singh, ``{{Invited Paper: Challenges and Opportunities in Enabling Secure {5G} Positioning}},'' in \emph{2023 15th International Conference on COMmunication Systems \& NETworkS (COMSNETS)}, 2023, pp. 498--504.

\bibitem{10127621}
F.~Morselli, S.~Modarres~Razavi, M.~Z. Win, and A.~Conti, ``{{Soft Information-Based Localization for {5G} Networks and Beyond}},'' \emph{IEEE Transactions on Wireless Communications}, vol.~22, no.~12, pp. 9923--9938, 2023.

\bibitem{9348766}
Y.~Li, Z.~Zhang, L.~Wu, J.~Dang, and P.~Liu, ``{5G} {C}ommunication {S}ignal {B}ased {L}ocalization with a {S}ingle {B}ase {S}tation,'' in \emph{2020 IEEE 92nd Vehicular Technology Conference (VTC2020-Fall)}, 2020, pp. 1--5.

\bibitem{3D}
\BIBentryALTinterwordspacing
B.~Sun, B.~Tan, W.~Wang, and E.~S. Lohan, ``A {C}omparative {S}tudy of {3D UE} {P}ositioning in {5G} {N}ew {R}adio with a {S}ingle {S}tation,'' \emph{Sensors}, vol.~21, no.~4, 2021. [Online]. Available: \url{https://www.mdpi.com/1424-8220/21/4/1178}
\BIBentrySTDinterwordspacing

\bibitem{9441655}
Y.~Liu, Z.~Gong, Z.~Zhang, L.~Wu, R.~Gui, J.~Dang, and B.~Zhu, ``High-{P}recision {S}ingle {B}ase {S}tation {L}ocalization {A}ssisted by {B}eamforming,'' in \emph{2021 2nd Information Communication Technologies Conference (ICTC)}, 2021, pp. 178--183.

\bibitem{9886342}
L.~Xie, Y.~Zhang, Y.~Wang, W.~Nie, and M.~Zhou, ``{{Multipath Assisted Single Base Station Positioning for NLOS Environment}},'' in \emph{2022 IEEE International Symposium on Antennas and Propagation and USNC-URSI Radio Science Meeting (AP-S/URSI)}, 2022, pp. 307--308.

\bibitem{8514133}
B.~Rares, C.~Codau, A.~Pastrav, T.~Palade, H.~Hedesiu, B.~Balauta, and E.~Puschita, ``{{Experimental Evaluation of AoA Algorithms using NI USRP Software Defined Radios}},'' in \emph{2018 17th RoEduNet Conference: Networking in Education and Research (RoEduNet)}, 2018, pp. 1--6.

\bibitem{LTEtb}
A.~Blanco, N.~Ludant, P.~J. Mateo, Z.~Shi, Y.~Wang, and J.~Widmer, ``{{Performance Evaluation of Single Base Station {ToA-AoA} Localization in an {LTE} Testbed}},'' in \emph{2019 IEEE 30th Annual International Symposium on Personal, Indoor and Mobile Radio Communications (PIMRC)}, 2019, pp. 1--6.

\bibitem{9880817}
D.~Li, X.~Chu, L.~Wang, Z.~Lu, S.~Zhou, and X.~Wen, ``{{Performance {E}valuation of {E-CID} based {P}ositioning on {OAI 5G-NR T}estbed}},'' in \emph{2022 IEEE/CIC International Conference on Communications in China (ICCC)}, 2022, pp. 832--837.

\bibitem{9841230}
Y.~Ge, H.~Chen, F.~Jiang, M.~Zhu, H.~Khosravi, S.~Lindberg, H.~Herbertsson, O.~Eriksson, O.~Brunnegård, B.-E. Olsson, P.~Hammarberg, F.~Tufvesson, L.~Svensson, and H.~Wymeersch, ``{{Experimental Validation of Single Base Station 5G mm Wave Positioning: Initial Findings}},'' in \emph{2022 25th International Conference on Information Fusion (FUSION)}, 2022, pp. 1--8.

\bibitem{3gppsrs}
3GPP, ``3{GPP TS} 38.211: {NR} {P}hysical channels and modulation ({R}elease 18),'' \emph{3rd {G}eneration {P}artnership {P}roject, {T}echnical {S}pecification {G}roup {R}adio {A}ccess {N}etwork}, 2024.

\bibitem{grove_barometer}
\BIBentryALTinterwordspacing
(2024) {{Grove - Barometer (High-Accuracy)}}. [Online]. Available: \url{https://wiki.seeedstudio.com/Grove-Barometer-High-Accuracy/}
\BIBentrySTDinterwordspacing

\bibitem{Xhafa2023}
A.~Xhafa, F.~Fabra, J.~A. López-Salcedo, G.~Seco-Granados, and I.~Lapin, ``{{Pathway to Coherent Phase Acquisition in Multi-Channel USRP SDRs for Direction of Arrival Estimation}},'' in \emph{2023 Tenth International Conference on Software Defined Systems (SDS)}, 2023, pp. 1--8.

\bibitem{me}
\BIBentryALTinterwordspacing
T.~Spanos, F.~Fabra, J.~A. L{\'o}pez-Salcedo, G.~Seco-Granados, N.~Kanistras, I.~Lapin, and V.~Paliouras, ``Angle of {A}rrival {E}stimation {U}sing {SRS} in {5G} {NR} {U}plink {S}cenarios,'' in \emph{WIPHAL}, 2024. [Online]. Available: \url{https://api.semanticscholar.org/CorpusID:271431302}
\BIBentrySTDinterwordspacing

\bibitem{ElHajjar2007}
C.~E. Hajjar, ``{{Synchronization Algorithms for OFDM Systems (IEEE802.11a, DVB-T): Analysis, Simulations, Optimization and Implementation Aspects}},'' Ph.D. dissertation, Dec. 2007, phD thesis.

\bibitem{cfo}
P.~K. Nishad and P.~Singh, ``{{Carrier frequency offset estimation in OFDM systems}},'' in \emph{2013 IEEE Conference on Information \& Communication Technologies}, 2013, pp. 885--889.

\bibitem{Akaike}
H.~Akaike, ``{{A new look at the statistical model identification}},'' \emph{IEEE Transactions on Automatic Control}, vol.~19, no.~6, pp. 716--723, 1974.

\bibitem{MUSIC}
R.~Schmidt, ``{{Multiple emitter location and signal parameter estimation}},'' \emph{IEEE Transactions on Antennas and Propagation}, vol.~34, no.~3, pp. 276--280, 1986.

\bibitem{ESPRIT}
R.~Roy and T.~Kailath, ``{{{ESPRIT}-estimation of signal parameters via rotational invariance techniques}},'' \emph{IEEE Transactions on Acoustics, Speech, and Signal Processing}, vol.~37, no.~7, pp. 984--995, 1989.

\bibitem{2DESPRIT}
A.-J. van~der Veen, M.~Vanderveen, and A.~Paulraj, ``{{Joint angle and delay estimation using shift-invariance techniques}},'' \emph{IEEE Transactions on Signal Processing}, vol.~46, no.~2, pp. 405--418, 1998.

\bibitem{ROOT-MUSIC}
A.~Barabell, ``{{Improving the resolution performance of eigenstructure-based direction-finding algorithms}},'' in \emph{ICASSP '83. IEEE International Conference on Acoustics, Speech, and Signal Processing}, vol.~8, 1983, pp. 336--339.

\bibitem{MVDR}
J.~Capon, ``{{High-resolution frequency-wavenumber spectrum analysis}},'' \emph{Proceedings of the IEEE}, vol.~57, no.~8, pp. 1408--1418, 1969.

\bibitem{lcmv}
O.~Frost, ``{{An algorithm for linearly constrained adaptive array processing}},'' \emph{Proceedings of the IEEE}, vol.~60, no.~8, pp. 926--935, 1972.

\end{thebibliography}

\end{document}